\documentclass[twocolumn,preprintnumbers,amsmath,aps]{revtex4}
\usepackage{graphicx}
\usepackage{dcolumn}
\usepackage{bm}
\usepackage{color}
\usepackage{amssymb}
\usepackage{array}
\begin{document}
\preprint{{\it preprint version}}
\title{Quantum conductance of silicon-doped carbon wire nanojunctions}
\author{D. Szcz{\c{e}}{\'s}niak$^{1,2}$}\email{d.szczesniak@ajd.czest.pl}
\author{A. Khater$^{1}$}
\author{Z. B{\c{a}}k$^{2}$}
\author{R. Szcz{\c{e}}{\'s}niak${^3}$}
\author{M. Abou Ghantous$^{4}$}
\affiliation{$^1$Institute for Molecules and Materials UMR 6283, University of Maine, Ave. Olivier Messiaen, 72085 Le Mans, France}
\affiliation{$^2$Institute of Physics, Jan D{\l}ugosz University in Cz{\c{e}}stochowa, Ave. Armii Krajowej 13/15, 42200 Cz{\c{e}}stochowa, Poland}
\affiliation{$^3$Institute of Physics, Cz{\c{e}}stochowa University of Technology, Ave. Armii Krajowej 19, 42200 Cz{\c{e}}stochowa, Poland}
\affiliation{$^4$Department of Physics, Texas A\&M University, Education City, PO Box 23874 Doha, Qatar}
\date{\today} 
\begin{abstract}

The unknown quantum electronic conductance across nanojunctions made of silicon-doped carbon wires between carbon leads is investigated. This is done by an appropriate generalization of the phase field matching theory for the multi-scattering processes of the electronic excitations at the nanojunction, and the use of the tight-binding method. Our calculations of the electronic band structures for carbon, silicon and diatomic silicon carbide, are matched with the available corresponding density functional theory results to optimize the required tight-binding parameters. The silicon and carbon atoms are treated on the same footing by characterizing each with their corresponding orbitals.  Several types of nanojunctions are analyzed to sample their behavior under different atomic configurations. We calculate for each nanojunction the individual contributions to the quantum conductance for the propagating $\sigma$, $\pi$, and $\sigma^{*}$ electrons incident from the carbon leads. The calculated results show a number of remarkable features, which include the influence of the ordered periodic configurations of silicon-carbon pairs, and the suppression of the quantum conductance due to minimum substitutional disorder and to artificially organized symmetry on these nanojunctions. Our results also demonstrate that the phase field matching theory is an efficient tool to treat the quantum conductance of complex molecular nanojunctions.

\end{abstract}
\maketitle
\noindent{\bf PACS numbers:} 85.35.-p, 73.63.Nm, 31.15.xf\\
\noindent{\bf Keywords:} nanoelectronics, quantum wires, electronic transport, finite-difference methods


\section{Introduction}

The quantitative analysis of the electronic quantum transport in nanostructures is essential for the development of nanoelectronic devices \cite{agrait}. The monatomic \textit{linear} carbon wire (MLCW) systems are expected in this context to have potentially interesting technological applications, in particular as connecting junction elements between the larger device components \cite{nitzan}. In this respect the electronic quantum transport properties are the key features of such wire nanojunctions \cite{wan}.

Carbon exists in nature under a wide range of allotropic forms, as the two-dimensional graphene \cite{geim}, the cage fullerenes \cite{kroto}, and the quasi-one-dimensional carbon nanotubes \cite{iijima}. These forms exhibit exceptional physical properties and can be considered as promising components for future nanodevices \cite{mceuen}. The discovery of monatomic linear carbon wires (MLCW), \cite{heath}, \cite{lagow}, \cite{derycke}, \cite{troiani}, \cite{zhao}, \cite{yuzvinsky}, \cite{jin} turns the attention to another intriguing carbon allotropic form. In the experiment conducted recently by Jin \textit{et al.} \cite{jin} the MLCW was produced by directly removing carbon atoms row by row from the graphene sheets, leading to a relatively stable freestanding nanostructure.

At present the available experimental data do not provide essential knowledge about the electronic properties of the MLCW systems, and only theoretical studies shed some light on these properties. Furthermore, although the MLCW systems were investigated for a long time from the theoretical point of view \cite{kertesz1}, \cite{kertesz2}, \cite{karpfen}, \cite{teramae}, \cite{springborg1}, \cite{rice}, \cite{springborg2}, \cite{watts}, \cite{xu}, \cite{lou}, \cite{jones}, \cite{fuentealba}, their interest was not highlighted until recently due to the open attention paid to other carbon allotropic forms. It has been shown in particular that from the structural point of view the MLCW can form either as cumulene wires (interatomic double bonds) or polyyne wires (alternating interatomic single and triple bonds) \cite{jin}, \cite{karpfen}, \cite{springborg1}, \cite{abdurahman}, \cite{cahangirov}. However, there is no straightforward answer which of these two structures is the favorable one; experimental studies do not give a satisfactory answer, and the theoretical calculations yield provisions which depend on the applied computational methods. The density functional theory (DFT) calculations predict double bond structures \cite{tongay1}, \cite{bylaska}, whereas the \textit{ab-initio} Hartree-Fock (HF) results favor the alternating bond systems \cite{kertesz1}, \cite{kertesz2}, \cite{karpfen}, \cite{teramae}, \cite{abdurahman}. This situation arises from the fact that the DFT tends to underestimate bond alternation (second order Jahn-Teller effect), while the HF overestimates it \cite{abdurahman}.

More recently, first-principle calculations have indicated \cite{zhang}, that both structures are stable and present mechanical characteristics of a purely one-dimensional nanomaterial. Moreover, on the basis of the first-principle calculations, \cite{zhang}, \cite{lang1}, \cite{lang2}, \cite{larade}, \cite{tongay2}, \cite{senger}, \cite{crljien}, \cite{okano}, \cite{chen1}, \cite{wang1},  \cite{song},\cite{zhang1}, the cumulene MLCW wires are expected to be almost perfect conductors, even better than linear gold wires \cite{tongay1}, while the corresponding polyyne wires are semiconducting \cite{song}. It is also worth noting that the MLCW cumulene system may exhibit conductance oscillations with the even and odd numbers of the wire atoms \cite{cahangirov}, \cite{zhang1}.

In the present work we consider in particular the problem of the electronic quantum transport across molecular nanojunctions made up of silicon-doped carbon wires, prepared in ordered or substitutionally disordered configurations as in the schematic representation of Figure 1, where the nanojunctions are between pure MLCW wire leads. This problem has not been considered previously and is still unsolved to our knowledge. The interest in the quantum transport of such nanojunctions arises from the fact that the chemical defects or substitutional disorder may have a significant impact on their transport properties \cite{ke}. Chemical impurities doping the nanojunction may even allow the control of the transport for such nanostructures \cite{nozaki}. The properties of the nanoelectronic device and its functionality may hence be greatly affected or even built on such ordered and disordered configurations. The interest in silicon carbide, furthermore,  stems from the fact that it is considered a good substrate material for the growth of graphene \cite{strupinski}, and may produce interesting effects in its interactions with Si or C \cite{wang2}.

The electrons which contribute to transport present characteristic wavelengths comparable to the size of molecular  nanojunctions, leading to quantum coherent effects. The transport properties of a given nanojunction are then described in terms of the Landauer-B{\"u}ttiker theory \cite{landauer}, \cite{buttiker}, which relate transmission scattering to quantum conductance. Several approaches have been developed in order to calculate the scattering transmission and reflection cross sections in nanostructures, where the most popular are based on first-principle calculations \cite{zwierzycki}, \cite{pauly} and semi-empirical methods using the nonequilibrium Green's function formalism \cite{caroli}, \cite{deretzis}.

In the present work we investigate the electronic scattering processes on the basis of the phase field matching theory (PFMT) \cite{khater1}, \cite{szczesniak}, originally developed for the scattering of phonons and magnons in nanostructures \cite{khater2}, \cite{khater3}, \cite{tigrine}, \cite{virlouvet}, \cite{fellay}. Our theoretical method is based on the appropriate phase matching of the Bloch states of ideal leads to the local states in the scattering region. In this approach the electronic properties of the system are described in the framework of the tight-binding formalism (TB) which is widely exploited for electronic transport calculations \cite{szczesniak}, \cite{mardaani}, \cite{rabani}, \cite{wu}, \cite{chen2}, and for  simulating the STM images of nanostructures \cite{hands}, \cite{delga}. In particular, we employ the appropriate Slater-Koster \cite{slater} type Hamiltonian parameters calculated on the basis of the Harrison tight-binding theory (HTBT) \cite{harrison}. The PFMT method, which is formally equivalent to the method of nonequilibrium Green's functions \cite{zhang5}, can be considered consequently as a transparent and efficient mathematical tool for the calculation of the electronic quantum transport properties for a wide range of molecular sized nanojunction systems.

The present paper is organized in the following manner. In section \ref{model}, we give the detailed discussion of the theoretical PFMT formalism. Our numerical results, which incorporate propagating and evanescent electronic states, are presented per individual lead modes in section \ref{results}. Also presented are the total conductance spectra. They are compared with results based on first-principle calculations when available. Finally the discussion and conclusions are given in section \ref{conclusions}. Appropriate appendices which supplement the theoretical model are presented in annex.

\section{Theoretical framework}
\label{model}

\subsection{Theoretical model and propagating states}
\label{model dynamics}
The schematic representation of the system under study with an arbitrary nanojunction region is presented in Figure 1. With reference to the Landauer-B{\"u}ttiker theory for the analysis of the electronic scattering processes \cite{landauer}, \cite{buttiker}, this system is divided into three main parts, namely the finite silicon-doped carbon wire nanojunction region, made up of a given composition of carbon (black) and silicon (orange) atoms, and two other regions to the  right and left of the nanojunction which are semi-infinite quasi one-dimensional carbon leads. Moreover, for the purpose of the quantum conductance calculations, the so-called irreducible region and the matching domains are depicted (see section \ref{pfmt} for more details). Figure 1 is used throughout section \ref{model} as a graphical reference for the analytical discussion.

The system presented in Figure 1 is described by the general tight-binding Hamiltonian block matrix
\begin{equation}
\label{eq1}
\mathbf{H}=\left[ \begin{array}{c c c c c}
\ddots & \cdots & 0 & 0 &  \\
\vdots & \mathbf{E}_{N-1,N-1} & \mathbf{H}_{N,N-1}^{\dagger} & 0 & 0 \\
0 &\mathbf{H}_{N,N-1} & \mathbf{E}_{N,N} & \mathbf{H}_{N+1,N}^{\dagger} & 0 \\
0 & 0 & \mathbf{H}_{N+1,N} & \mathbf{E}_{N+1,N+1} & \vdots \\
  & 0 & 0 & \cdots & \ddots
\end{array} \right].
\end{equation}
This is defined in general for a system of $N_x$ inequivalent atoms per unit cell, where $N_l$ denotes the number of the basis orbitals per atomic site, assuming spin degeneracy. In equation (\ref{eq1}), $\mathbf{E}_{i,j}$ denote the on-diagonal matrices composed of both diagonal $\varepsilon_{l}^{n,\alpha}$ and off-diagonal $h_{l,l',m}^{n,n',\beta}$ elements for a selected unit cell. In contrast the $\mathbf{H}_{i,j}$ matrices contain only off-diagonal elements for the interactions between different unit cells. The index $\alpha$ identifies the atom type, C or Si, on the $n$th site in a unit cell. Each diagonal element is characterized by the lower index $l$ for the angular momentum state. The off-diagonal elements $h_{l,l',m}^{n,n',\beta}$ describe the $m$-type bond, ($m=\sigma,$ $\pi$), between $l$ and $l'$ nearest-neighbor states. The index $\beta$ identifies the types of interacting neighbors, C-C, Si-Si, or Si-C.

\begin{figure}[ht]
\centering
\includegraphics[width=\columnwidth]{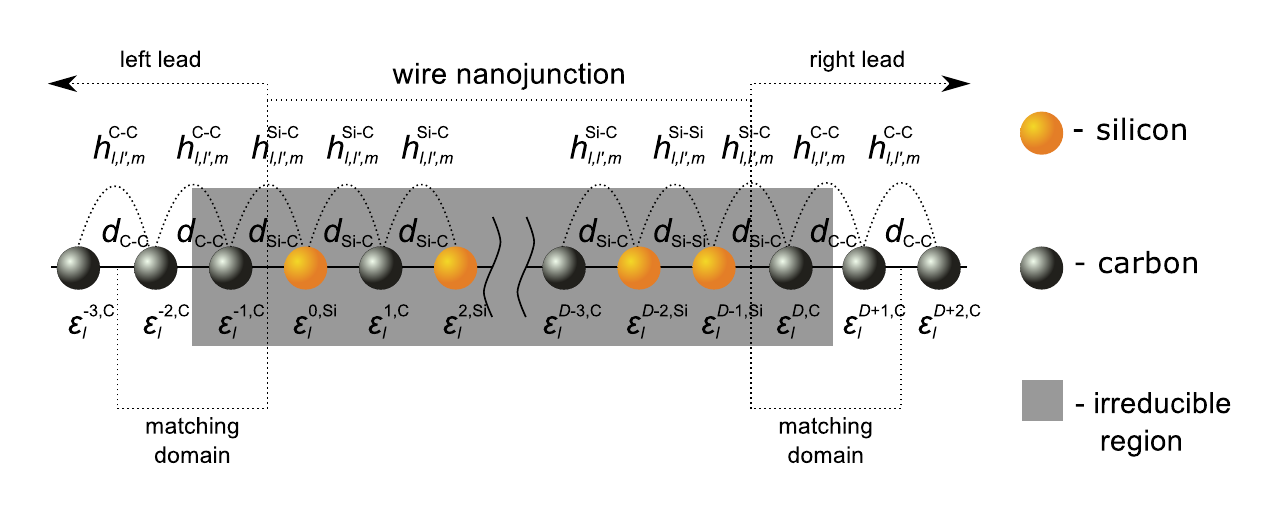}
\caption{Schematic representation of the finite silicon-doped carbon wire nanojunction between two semi-infinite quasi one-dimensional carbon leads. The irreducible region and matching domains are distinguished (please see section \ref{pfmt} for more details). The binding energies for a given atomic site and the coupling terms between neighbor atoms with corresponding interatomic distances are depicted. The $n$ and $n'$ indices for the coupling parameters are dropped for simplicity.}
\label{fig01}
\end{figure}

The $h_{l,l',m}^{n,n',\beta}$ elements are consistent with the Slater-Koster convention \cite{slater}, and may be expressed in the framework of the Harrison's tight-binding theory (HTBT) \cite{harrison}, by
\begin{equation}
\label{eq2}
h_{l,l',m}^{n,n',\beta} = \eta_{l,l',m} \frac{\hslash^2}{m_e d^{2}_{\beta}},
\end{equation}
where $\eta_{l,l',m}$ are the dimensionless Harrison coefficients, $m_e$ the electron mass in vacuum, and $d_{\beta}$ the interatomic distance for interacting neighbors. Explicit forms of the $\mathbf{E}_{i,j}$ and $\mathbf{H}_{i,j}$ matrices are given in appendix A. The tight-binding parameters scheme are illustrated in Figure 1; note however that the $n$ and $n'$ indices for the coupling parameters are dropped for simplicity in this figure.

In our calculations the single-particle electronic wave functions are expanded in the orthonormal basis of local atomic wave functions $\phi_l ({\bf r})$ as follows
\begin{equation}
\label{eq3}
\Psi (\mathbf{r}, \mathbf{k}) =  \sum_{l, n, N} c_l (\mathbf{r}_{n}-\mathbf{R}_{N}, \mathbf{k}) \phi_l (\mathbf{r}-\mathbf{R}_{N}, \mathbf{k}).
\end{equation}
In equation (\ref{eq3}), $\mathbf{k}$ is the real wave vector, $\mathbf{R}_{N}$ the position vector of the selected unit cell, and $\mathbf{r}_{n}$ the position vector of the $n$th atom in the selected unit cell. For the ideal leads, the wavefunction coefficients $c_l (\mathbf{r}_{n}-\mathbf{R}_{N}, \mathbf{k})$ are characterized under the Bloch-Floquet theorem in consecutive unit cells by the following phase relation
\begin{equation}
\label{eq4}
c_l ( \mathbf{r}_{n}-\mathbf{R}_{N+1}, \mathbf{k} ) = z c_l ( \mathbf{r}_{n}-\mathbf{R}_{N}, \mathbf{k} ),
\end{equation}
where $z$ is the phase factor
\begin{equation}
\label{eq5}
z_{\pm}=e^{\pm i\mathbf{k}\mathbf{R}_{N}},
\end{equation}
which correspond here to waves propagating to the right (+) or to the left (-).

The electronic equations of motion for a leads unit cell, independent of $N$, may be expressed in square matrix form, with an orthonormal minimal basis set of local wavefunctions, as
\begin{equation}
\label{eq6}
(E\mathbf{I} - \mathbf{M}_d)\times\mathbf{c}( \mathbf{k}, E )=0.
\end{equation}
$E$ stands for the electron eigenvalues, $\mathbf{I}$ is the  identity matrix, while the dynamical matrix $\mathbf{M}_d$ contains the Hamiltonian matrix elements and the $z$ phase factors. The $\mathbf{c} (\mathbf{k},E)$ is the $N_x \times N_l$ size vector defined as
\begin{equation}
\label{eq7}
\mathbf{c} (\mathbf{k}, E)=\left[ \begin{array}{c} c_s (\mathbf{r}_1, \mathbf{k}, E) \\ c_{p_x} (\mathbf{r}_1, \mathbf{k}, E) \\ \vdots \\ c_{p_y} (\mathbf{r}_n, \mathbf{k}, E)\\c_{p_z} (\mathbf{r}_n, \mathbf{k}, E) \end{array} \right] \equiv \left[ \begin{array}{c} \mathbf{c}_l (\mathbf{r}_1, \mathbf{k}, E) \\ \vdots \\ \mathbf{c}_{l} (\mathbf{r}_n, \mathbf{k}, E)\end{array} \right].
\end{equation}
Equation (\ref{eq6}) gives the $N_x \times N_l$ eigenvalues with corresponding eigenvectors which determine the electronic structure of the lead system, where $l$ under the vector $\mathbf{c}_l$ corresponds to the $N_l=4$ orbitals $s, p_x, p_y, p_z$. Note that the choice of an orthonormal minimal basis set of local wavefunctions may result initially in an inadequate description of the considered electronic eigenvalues. However, as can be seen later, the proper choice of the TB on-site energies and coupling terms allows us to to obtain agreement with the DFT results. This is a systematic procedure in our calculations.
\subsection{Evanescent states}
The complete description of the electronic states on the ideal leads requires a full understanding of the propagating and of the evanescent electronic states on the leads. This arises because the silicon-doped nanojunction breaks the perfect periodicity of the infinite leads, and forbids a formulation of the problem only in terms of the pure Bloch states as given in equation (\ref{eq5}). Depending on the complexity of a given electronic state, it follows that the evanescent waves may be defined by the phase factors for a purely imaginary wave vectors $\mathbf{k}=i \boldsymbol{\kappa}$ such that
\begin{equation}
\label{eq8}
z=z_{\pm}=e^{\mp \boldsymbol{\kappa} \mathbf{r}_n},
\end{equation}
or for complex wave vectors $\mathbf{k}=\boldsymbol{\kappa}_1+i \boldsymbol{\kappa}_2$ such that
\begin{equation}
\label{eq9}
z=z_{\pm}=e^{\mp (i \boldsymbol{\kappa}_1-\boldsymbol{\kappa}_2) \mathbf{r}_n}.
\end{equation}
The phase factors of equations (\ref{eq8}) and (\ref{eq9}) correspond to pairs of hermitian evanescent and divergent solutions on the leads. Only the evanescent states are physically considered, where the spatial evanescence occurs to the right and left away from the nanojunction localized states. Note that the $l$-type evanescent state corresponds to energies beyond the propagating band structure for this state.

The functional behavior of $z(E)$ for the propagating and evanescent states on the leads may be obtained by various techniques. An elegant method presented previously for phonon and magnon excitations \cite{fellay}, is adapted here for the electrons. It is described on the basis of equations (\ref{eq4}) and (\ref{eq6}) by the generalized eigenvalue problem for $z$
\begin{eqnarray}
\label{eq10}
\nonumber
&&\left[\left[ \begin{array}{c c}
E\mathbf{I}-\mathbf{E}_{N,N} & \mathbf{H}_{N,N-1} \\
\mathbf{I} & 0
\end{array}\right] - z
\left[ \begin{array}{c c}
-\mathbf{H}^{\dagger}_{N,N-1} & 0 \\
0 & \mathbf{I}
\end{array}\right]
\right]\\
&&\times \left[ \begin{array}{c}
\mathbf{c} ( \mathbf{R}_N, z, E ) \\
\mathbf{c} ( \mathbf{R}_{N-1}, z, E )
\end{array} \right]=0.
\end{eqnarray}
Equation (\ref{eq10}) gives the 2$N_x N_l$ eigenvalues as an ensemble of $N_x N_l$ pairs of $z$ and $z^{-1}$. Only solutions with $|z|=1$ (propagating waves) and $|z|<1$ (evanescent waves) are retained as a physical ones. In equation (\ref{eq10}) $\mathbf{k}$ is then replaced by the appropriate energy $E$ variable. Furthermore, for systems with more than one atom per unit cell, the matrices $\mathbf{H}_{N,N-1}$ and $\mathbf{H}^{\dagger}_{N,N-1}$ in this procedure are singular. In order to obtain the physical solutions, the eigenvalue problem of equation (\ref{eq10}) is reduced from the 2$N_x N_l$ size problem to the appropriate $2N_l$ one, by using the partitioning technique (\ref{appendixb}).

\subsection{Phase field matching theory}
\label{pfmt}

Consider next the scattering problem at the nanojunction. An electron incident along the leads has a given energy $E$ and wave vector $\mathbf{k}$, where $E = E_{\gamma}(\mathbf{k})$ denotes the available dispersion curves for the $\gamma$  = 1, 2,..., $\Gamma$ propagating eigenmodes, where $\Gamma$ corresponds to the total number of allowed solutions for the eigenvalue problem of phase factors in equation (\ref{eq10}). In any given energy interval, however, these may be evanescent or  propagating eignemodes, and together constitute a complete set of available channels necessary for the scattering analysis.

The irreducible domain of atomic sites for the scattering problem includes the nanojunction domain itself, ($N\in[0, D-1]$), and the atomic sites on the left and right leads which interact with the nanojunction, as in Figure 1. This constitutes a necessary and sufficient region for our considerations, {\it i.e.} any supplementary atoms from the leads included in the calculations do not change the final results. The scattering at the boundary yields then the coherent reflected and transmitted fields, and in order to calculate these, we establish the \textit{system} of equations of motion for the atomic sites ($N\in[-1, D]$) of the irreducible nanojunction domain. This procedure leads to the general matrix equation
\begin{equation}
\label{eq11}
\mathbf{M}_{nano}\times\mathbf{V}=0.
\end{equation}
$\mathbf{M}_{nano}$ is a $(D+2)\times(D+4)$ matrix composed of the block matrices $(E\mathbf{I} - \mathbf{E}_{N,N} - \mathbf{H}_{N,N-1} - \mathbf{H}_{N,N-1}^{\dagger})$, and the state vector $\mathbf{V}$ of dimension $D+4$ is given as
\begin{equation}
\label{eq12}
\mathbf{V} = \left[ \begin{array}{c}
\mathbf{c}_{l} ( \mathbf{r}_{1}-\mathbf{R}_{-2}, E)\\
\vdots \\
\mathbf{c}_{l} ( \mathbf{r}_{n}-\mathbf{R}_{-2}, E )\\
\vdots \\
\vdots \\
\mathbf{c}_{l} ( \mathbf{r}_{1}-\mathbf{R}_{D+1}, E ) \\
\vdots \\
\mathbf{c}_{l} ( \mathbf{r}_{n}-\mathbf{R}_{D+1}, E )\\
\end{array} \right].
\end{equation}
Since the number of unknown coefficients in equation (\ref{eq11}) is always greater than the number of equations, such a set of equations cannot be solved directly.

Assuming that the incoming electron wave propagates from left to right in the eigenmode $\gamma$ over the interval of energies $E=E_\gamma$, the field coefficients on the left and right sides of the irreducible nanojunction domain may be written as
\begin{eqnarray}
\label{eq13}
\nonumber
\mathbf{c}_{l}^{L} (\mathbf{r}_{n}-\mathbf{R}_N, z_\gamma, E_\gamma) &=& \mathbf{c}_{l} (\mathbf{r}_{n}, z_\gamma, E_\gamma) z_\gamma^{-N}\\ \nonumber
&+& \sum_{\gamma'}^{\Gamma}  \mathbf{c}_{l} (\mathbf{r}_{n}, z_{\gamma'}, E_\gamma) z_{\gamma'}^{N} r_{\gamma,\gamma'}( E_\gamma) \hspace{0.2cm}\\ &\textrm{for}& \hspace{0.2cm} N \leqslant -1,
\end{eqnarray}
\begin{eqnarray}
\label{eq14}
\nonumber
\mathbf{c}_{l}^{R} (\mathbf{r}_{n}-\mathbf{R}_N, z_\gamma, E_\gamma) &=& \sum_{\gamma'}^{\Gamma} \mathbf{c}_{l} (\mathbf{r}_{n},z_{\gamma'}, E_\gamma) z_{\gamma'}^{N} t_{\gamma,\gamma'} (E_\gamma)\hspace{0.2cm}\\ &\textrm{for}& \hspace{0.2cm} N \geqslant D,
\end{eqnarray}
where $\gamma'\in \Gamma$ is an arbitrary channel into which the incident electron wave scatters, and $\mathbf{c}_{l} (\mathbf{r}_{n}, z_\gamma, E_\gamma)$ denotes the the eigenvector of the lead dynamical matrix of equation (\ref{eq6}) for the inequivalent site $n$, at $z_{\gamma}$ and $E_\gamma$. The terms $r_{\gamma,\gamma'}$ and $t_{\gamma,\gamma'}$ denote the scattering amplitudes for respectively backscattering and transmission from the $\gamma$ into the $\gamma'$ eigenmodes, and constitute the basis of the Hilbert space which describes the reflection and transmission processes.

Equations (\ref{eq13}) and (\ref{eq14}) are next used to transform the $(D+2) \times (D+4)$ matrix of the system of  equations of motion, equation (\ref{eq11}), into an inhomogeneous $(D+2) \times (D+2)$ matrix for the scattering problem. This procedure leads to the new form of the vector

\begin{eqnarray}
\label{eq15}
\mathbf{V}&=&
\left[ \begin{array}{c c c c c c c}
z^2 & 0 & \cdots & \cdots & \cdots & \cdots & 0 \\
z & 0 &  &  &  &  &\vdots \\
0 & 1 &  &  &  &  & \vdots \\
\vdots & & \ddots &  &  &  & \vdots\\
\vdots &  &  & 1 &  &  & \vdots\\
\vdots &  &  &  & \ddots &  & \vdots \\
\vdots &  &  &  &  & 1 & 0 \\
\vdots &  &  &  &  & 0 & z \\
0 & \cdots & \cdots & \cdots & \cdots & 0 & z^2
\end{array} \right]\\ \nonumber
&\times&\left[ \begin{array}{c}
\mathbf{r}_{\gamma,\gamma'}\\
\mathbf{c}_{l} ( \mathbf{r}_{1}-\mathbf{R}_{0}, E_\gamma )\\
\vdots \\
\mathbf{c}_{l} ( \mathbf{r}_{n}-\mathbf{R}_{0}, E_\gamma )\\
\vdots \\
\vdots \\
\mathbf{c}_{l} ( \mathbf{r}_{1}-\mathbf{R}_{D-1}, E_\gamma ) \\
\vdots \\
\mathbf{c}_{l} ( \mathbf{r}_{n}-\mathbf{R}_{D-1}, E_\gamma ) \\
\mathbf{t}_{\gamma,\gamma'}\\
\end{array} \right]
+ \left[ \begin{array}{c}
\mathbf{c}_{l} (\mathbf{r}_{1}, z_\gamma, E_\gamma ) z^{-2}_\gamma \\
\vdots \\
\mathbf{c}_{l} (\mathbf{r}_{n}, z_\gamma, E_\gamma ) z^{-2}_\gamma \\
\mathbf{c}_{l} (\mathbf{r}_{1}, z_\gamma, E_\gamma ) z^{-1}_\gamma \\
\vdots \\
\mathbf{c}_{l} (\mathbf{r}_{n}, z_\gamma, E_\gamma ) z^{-1}_\gamma \\
0 \\
\vdots \\
\vdots \\
0\\
\end{array} \right].
\end{eqnarray}
The rectangular sparse matrix in equation (\ref{eq15}) has $(D+4) \times (D+2)$ size. The vectors $\mathbf{r}_{\gamma,\gamma'}$ and $\mathbf{t}_{\gamma,\gamma'}$  are column vectors of the backscattering and transmission Hilbert basis.

Substituting equation (\ref{eq15}) into equation (\ref{eq11}) yields an inhomogenous system of equations as follows
\begin{equation}
\label{eq16}
\mathbf{M}\times \left[ \begin{array}{c}
\mathbf{r}_{\gamma,\gamma'}\\
\mathbf{c}_{l} ( \mathbf{r}_{1}-\mathbf{R}_{0}, E_\gamma )\\
\vdots \\
\mathbf{c}_{l} ( \mathbf{r}_{n}-\mathbf{R}_{0}, E_\gamma )\\
\vdots \\
\vdots \\
\mathbf{c}_{l} ( \mathbf{r}_{1}-\mathbf{R}_{D-1}, E_\gamma ) \\
\vdots \\
\mathbf{c}_{l} ( \mathbf{r}_{n}-\mathbf{R}_{D-1}, E_\gamma ) \\
\mathbf{t}_{\gamma,\gamma'}\\
\end{array} \right]
= - \left[ \begin{array}{c}
\mathbf{M}^{in}_1\\
\mathbf{M}^{in}_2\\
0 \\
\vdots \\
0\\
\end{array} \right].
\end{equation}
In equation \ref{eq16}, $\mathbf{M}$ is the \textit{matched} $(D+2) \times (D+2)$ square matrix, and the vector of dimension $(D+2)$ which incorporates the $\mathbf{M}^{in}_{1}$ and $\mathbf{M}^{in}_{2}$ elements, regroups the inhomogeneous terms of the incident wave. The explicit forms of the $\mathbf{M}$ matrix elements and and $\mathbf{M}^{in}_{N}$ vectors are presented in \ref{appendixc}.

In practice, equation (\ref{eq16}) can be solved using standard numerical procedures, over the entire range of available electronic energies, yielding the coefficients $\mathbf{c}_{l}$ for the atomic sites on the nanojunction domain itself, and also the $\Gamma$ reflection $r_{\gamma,\gamma'}(E)$ and the $\Gamma$ transmission $t_{\gamma,\gamma'}(E)$ coefficients.

The reflection and transmission coefficients give respectively the reflection $R_{\gamma,\gamma'}(E)$ and transmission $T_{\gamma,\gamma'}(E)$ probabilities, by normalizing with respect to their group velocities $v_{\gamma}$ in order to obtain the unitarity of the scattering matrix, as
\begin{equation}
\label{eq17}
R_{\gamma,\gamma'}(E) = \frac{v_{\gamma'}}{v_\gamma} \left| r_{\gamma,\gamma'}(E) \right|^2,
\end{equation}
\begin{equation}
\label{eq18}
T_{\gamma,\gamma'}(E) = \frac{v_{\gamma'}}{v_\gamma} \left| t_{\gamma,\gamma'}(E) \right|^2,
\end{equation}
where $v_{\gamma}\equiv v_{\gamma}(E)$ denotes the group velocity of the incident electron wave in the eigenmode  $\gamma$. The group velocities are calculated by a straightforward procedure as in \ref{appendixd}. For the evanescent eigenmodes $v_{\gamma'}$ = 0. Although the evanescent eigenmodes do not contribute to the electronic transport, they are required for the complete description of the scattering processes.

Furthermore, using equations (\ref{eq17}) and (\ref{eq18}), the overall reflection probability, $R_{\gamma}(E)$, for an electron incident in the $\gamma$ eigenmode, and the total electronic reflection probability, $R(E)$, from all the eignemodes may be expressed respectively as
\begin{equation}
\label{eq20}
R_{\gamma}(E) = \sum_{\gamma'}^{\Gamma} R_{\gamma,\gamma'}(E) \hspace{0.2cm} \textrm{and} \hspace{0.2cm} R(E) = \sum_{\gamma}^{\Gamma} R_{\gamma}(E).
\end{equation}
Similarly, for the transmission probabilities, we may write the equivalent equations as
\begin{equation}
\label{eq21}
T_{\gamma}(E) = \sum_{\gamma'}^{\Gamma} T_{\gamma,\gamma'}(E) \hspace{0.2cm} \textrm{and} \hspace{0.2cm} T(E) = \sum_{\gamma}^{\Gamma} T_{\gamma}(E).
\end{equation}
The $T_{\gamma}(E)$ and $T(E)$ probabilities are very important for the electronic scattering processes since they correspond directly to the experimentally measurable observables. Likewise, the total transmission $T(E_\gamma)$ allows to calculate the overall electronic conductance. In this work we assume the zero bias limit and write the total conductance in the following way
\begin{equation}
\label{eq22}
G(E_F)=G_0 T(E_F).
\end{equation}
In equation (\ref{eq22}), $G_0$ is the conductance quantum and equals $2e^2/h$. Due to the Fermi-Dirac distribution the $G(E_F)$ is calculated at the Fermi level of the perfect leads band structure, since electrons only at this level give the important contribution to the electronic conductance. The Fermi energy can be determined using various methods, where in the present work $E_F$ is calculated the basis of the density of states calculations.

\section{Numerical results and discussion}
\label{results}

%
\subsection{The tight-binding model and basic electronic properties}
\label{electronicstructure}
In subsection \ref{electronicstructure} we present the results of our model calculations for the electronic structure of the carbon, silicon and silicon carbide wires under study. Our results are validated by comparison with the DFT calculations \cite{tongay1}, \cite{bekaroglu}, which allows us to establish unambiguously our choice of the tight-binding parameters for these systems.

In principle, we can develop our model calculations for the nanojunctions and their leads using any adequate type of orbitals; even a single orbital suffices to calculate the electronic quantum transport for carbon nanojunctions, Nozaki \textit{et al.} \cite{nozaki}. However, this approximation is inadequate for the silicon atoms. To treat both types of atoms on the same footing, we characterize hence the atoms by the electronic states 2$s$ and 2$p$ for carbon, and by the 3$s$ and 3$p$ for silicon. Such a scheme gives us four different orbitals, namely $s$, $p_x$, $p_y$ and $p_z$, for both types of atoms.

\begin{table}

\caption{The values of the tight-binding parameters, $\varepsilon_{l}^{n,\alpha}$ and $h_{l,l',m}^{n,n',\beta}$ (in eV), and Harrison's dimensionless coefficients, $\eta_{l,l',m}$, as proposed in this work, and compared with the original values by Harrison \cite{harrison}. Please note that the distance dependent $h_{l,l',\sigma}^{n,n',\beta}$ parameters are computed for the appropriate interatomic spacings $d_{\beta}$ (in \AA), tabulated below and assumed after \cite{tongay1} and \cite{bekaroglu}. In order to keep table transparent indices $n$ and $n'$ for $\varepsilon_{l}^{n,\alpha}$ and $h_{l,l',m}^{n,n',\beta}$ are ommited.}

\begin{tabular}{ c c c c c c c }

\hline

 & \multicolumn{3}{ c }{Harrison TB parameters} & \multicolumn{3}{ c }{Our TB parameters} \\ \hline

\begin{tabular}{ c }
$\alpha$ \\
$\varepsilon_{s}$\\
$\varepsilon_{p}$\\
\end{tabular}

 & \multicolumn{3}{ c }{\begin{tabular} { c c } C & Si \\

-19.38 & -14.79 \\

-11.07 & -7.59 \\

 \end{tabular}}

 & \multicolumn{3}{ c }{\begin{tabular} { c c } C & Si \\

-18.89 & -13.5 \\

-10.94 & -8.38

  \end{tabular}} \\ \hline

$\beta$ & C-C & Si-Si & Si-C & C-C &  Si-Si & Si-C \\

$\eta_{s,s,\sigma}$ & -1.32 & -1.32 & -1.32 & -0.93 & -1.48 & -1.11 \\

$\eta_{s,p,\sigma}$ & 1.42 & 1.42 & 1.42 & 0.94 & 1.19 & 0.95 \\

$\eta_{p,p,\sigma}$ & 2.22 & 2.22 & 2.22 & 1.03 & 1.18 & 0.99 \\

$\eta_{p,p,\pi}$ & -0.63 & -0.63 & -0.63 & -0.59 & -0.41 & -0.62 \\ \hline

$\beta$ & C-C & Si-Si & Si-C & C-C &  Si-Si & Si-C \\

$h_{s,s,\sigma}$ & -5.95 & -2.08 & -3.70 & -4.19 & -2.33 & -3.11 \\

$h_{s,p,\sigma}$ & 6.40 & 2.24 & 3.98 & 4.23 & 1.87 & 2.66 \\

$h_{p,p,\sigma}$ & 10.01 & 3.50 & 6.22 & 4.64 & 1.86 & 2.77 \\

$h_{p,p,\pi}$ & -2.84 & -0.99 & -1.77 & -2.66 & -0.65 & -1.74 \\ \hline

\begin{tabular} { c }
$\beta$\\
$d_{\beta}$
\end{tabular}

& \multicolumn{2}{ c }{\begin{tabular} { c } C-C \\
1.3
\end{tabular}}

& \multicolumn{2}{ c }{\begin{tabular} { c } Si-Si \\
2.2
\end{tabular}}

& \multicolumn{2}{ c }{\begin{tabular} { c } Si-C \\
1.649
\end{tabular}}\\ \hline
\end{tabular}
\end{table}

In the present work our TB parameters are effectively rescaled from the Harrison's data in order to match our model calculations for the electronic structure with those given by the DFT. The utilized TB parameters are presented in Table 1, in comparison with the values given by  Harrison. Note that the values of the on-site Hamiltonian matrix elements $\varepsilon_{p}^{n,\alpha}$ are identical for states $p_x$, $p_y$ and $p_z$. The off-diagonal distance dependent $h_{l,l',m}^{n,n',\beta}$  elements are calculated on the basis of equation (\ref{eq2}). For symmetry considerations, these latter are positive or negative, also $h_{s,p,\sigma}$ = $\eta_{s,p,\sigma}$ = 0 and $h_{p,p,\sigma}$ = $\eta_{p,p,\sigma}$ = 0, for $p_y$ and $p_z$, and $h_{p,p,\pi}$ = $h_{p_y,p_y,\pi}$ = $h_{p_z,p_z,\pi}$ = $h_{p_x,p_x,\pi}$ = 0 \cite{kaxiras}. Table 1 is supplemented for the reader by Figures 2 which give the dependence of the hopping integrals with distance as calculated in the present paper (continuous curves), in comparison with the Harrison's data (open symbols).

Figure 2 clearly indicates the fact that qualitatively both Harrison's and our rescaled coupling parameters for silicon, carbon and diatomic silicon carbide wires, present the same functional behavior, confirming the desired conservation of their physical character. However, most of the rescaled coupling parameters have somehow smaller values then those initially proposed by Harrison; this trend can be also traced in Table 1 for the onsite parameters. This difference  stems from the influence of the low-coordinated systems considered here, whereas the initial Harrison values are given to match tetrahedral phases \cite{harrison}. Another general observation can be made for the tight-binding parameters of the $\sigma$-type interactions (the  $h_{s,p,\sigma}$ and  $h_{p,p,\sigma}$ ones), which present much closer values over the considered interatomic distance range then in the case of the Harrison data.

\begin{figure}[ht]
\centering
\includegraphics[width=\columnwidth]{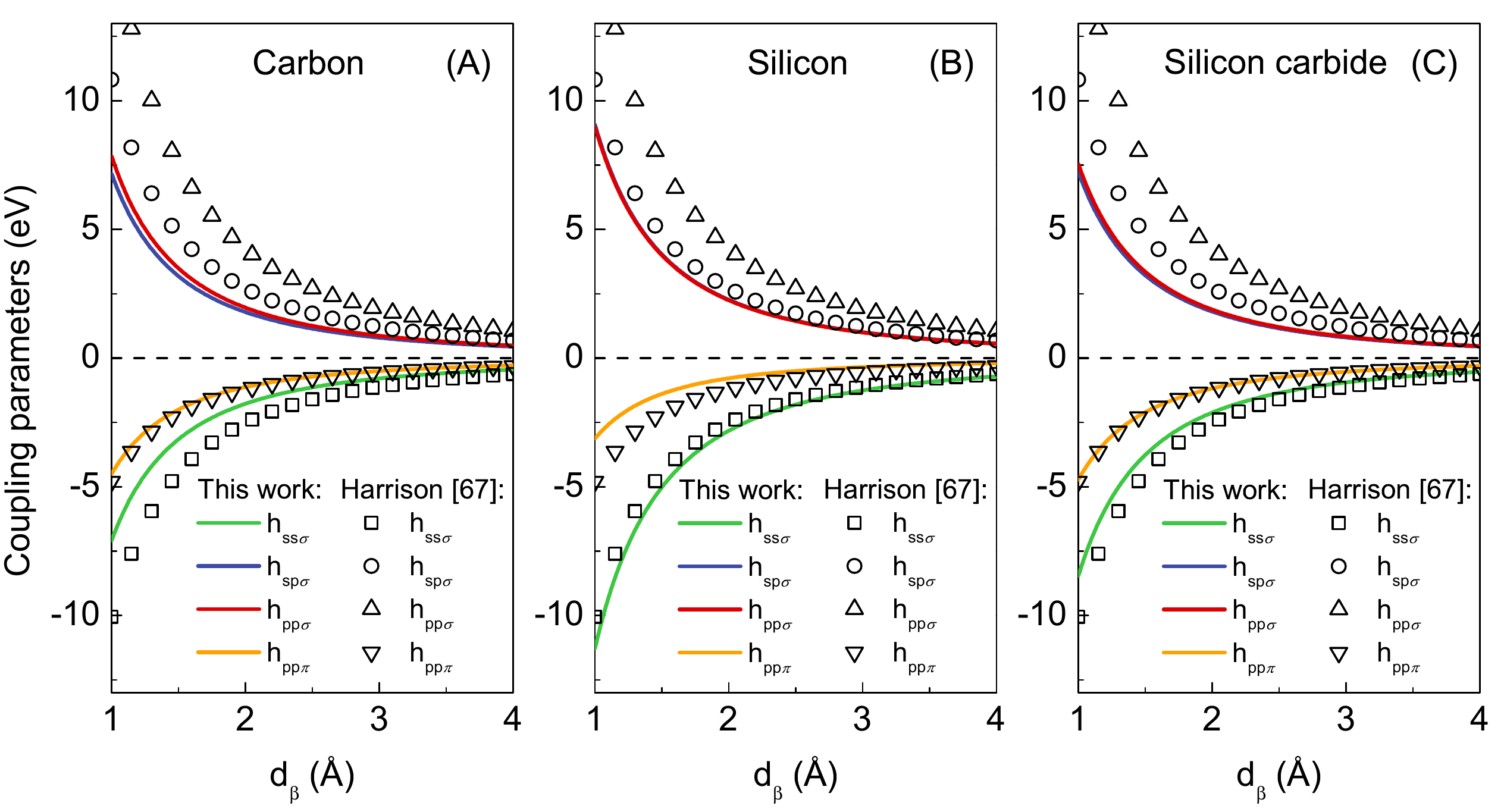}
\caption{The nearest-neighbor tight-binding coupling parameters with the interatomic distance. The curves represent our calculated TB results in comparison with those calculated using the Harrison parameters (squares, triangles, circles).}
\label{fig02}
\end{figure}

Our calculated electronic band structures for silicon, carbon and diatomic silicon carbide infinite wires (continuous curves) are presented in Figures 3, in comparison with the DFT results, \cite{tongay1}, \cite{bekaroglu}, as in the \textit{right hand} side of the figures. We note for the carbon and silicon structures that our TB parameters correctly reproduce the DFT results up to energies slightly above the Fermi level. Electronic branches in the regions of high-energies are in qualitative agreement. In the case of the diatomic silicon carbide structure some of the electronic states perfectly match the DFT results even for the high-energy domains. The \textit{left hand} side of Figures 3 compare our results (continuous curves) with those from the older TB values given by Harrison (open symbols); as is seen our TB parameters constitute the most optimal set for the electronic transport calculations, since their corresponding electronic band structures conform to the appropriate energy ranges highlighted by the DFT results, and what is even more important, correctly reproduce the Fermi level.

In Figures 3 (A) and (B) for silicon and carbon, the red and blue colors correspond respectively to the $\sigma$ and $\sigma^*$ bands. These arise from the $sp_x$ orbital hybrids, where the lowest lying bands are always occupied by two electrons. Bands marked by the red color have the $\pi$ character and are degenerate. Their origin in the $p_y$ and $p_z$ orbitals allows them to hold up to four electrons. In Figure 3 (C) for the diatomic silicon carbide, starting from the band structure minimum, consecutive bands have their origin in the following orbitals: carbon 3$s$ (red band), silicon 3$s$ (green band), carbon 3$p$ (blue and black bands), and silicon 3$p$ (orange and violet bands). The blue and orange colors for the silicon carbide electronic structure indicate two doubly degenerate $\pi$-type bands.

The metallic or insulating character of the considered atomic wires, following the Fermi level, is appropriate only when the wires are infinite. It is well known that this character can change for the case of finite size wires with a limited number of atoms, or due to the type and quality of the leads.

\subsection{Numerical characteristics for the carbon leads}

In general, the infinite carbon wires which are considered as the leads in our work, present electronic band structure characteristics which incorporate not only propagating, see Figure 3 (A), but also evanescent states. Both of these types of states, which are derivable from the generalized eigenvalue problem as presented in Eq.10, constitute a complete set over the allowed energies for the electrons incident along the leads, which can be further scattered at the considered nanojunction. This complete set of eigenstates is used as the basis for the numerical calculations of the quantum conductance presented in subsection \ref{transportproperties}.

\begin{figure}[ht]
\centering
\includegraphics[width=\columnwidth]{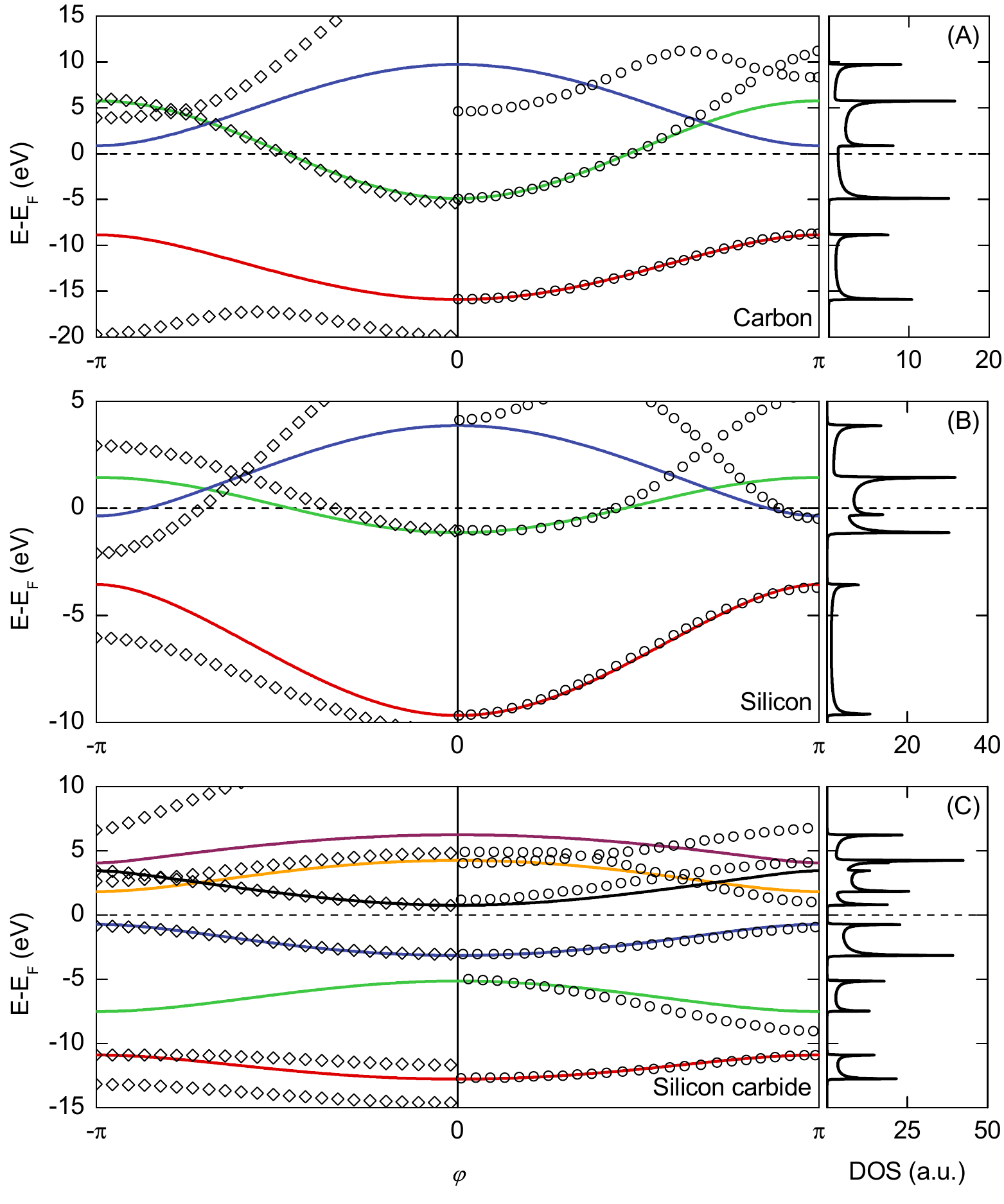}
\caption{Electronic structures of carbon (A), silicon (B) and diatomic silicon carbide (C) for infinite linear atomic wires, are presented over the first Brillouin zone $\varphi = kd \in [-\pi, \pi]$. Our calculated results (\textit{continuous curves}), represented by a color scheme (details in the text), are compared on the right hand side with the first-principle results (closed circles, $\varphi \in [0, \pi]$) \cite{tongay1}, \cite{bekaroglu}, and on the left hand side with results calculated using Harrison TB parameters, \cite{harrison} (diamonds, $\varphi \in [-\pi, 0]$). Our calculated Fermi levels are given as the zero reference energies, and the calculated electronic DOS in arbitrary units are presented in the right hand column.}
\label{fig02}
\end{figure}

Figure 4 (A) presents the three-dimensional representation of the solutions of Eq.10 as a set of generalized functionals $z(E)$ for the $\sigma$, $\sigma^*$ and $\pi$ electronic states of the carbon leads. As described by Eqs.5, 8 and 9, the eigenstates in Figure 4 (A) characterized by $|z|=1$, correspond to the propagating electronic waves described by the real wave vectors, whereas those by $|z|<1$ to the evanescent and divergent eigenstates for the complex wave vectors. Furthermore, for convenience the corresponding moduli of the complex $z$ factors are presented in Figure 4 (B). Note that the $|z|=1$ solutions may be grouped into pairs for the two directions of propagation linked by the time-reversal symmetry. Due to the fact that each of these two solutions provides the same information, we consider waves propagating only from left to right. However, this is not true for the $|z|<1$ solutions which are always considered for both left and right as spatially  evanescent. As can be seen in Figures 4, the generalized results for $\sigma$, $\sigma^*$ and $\pi$ states are represented by the same colors as the corresponding states in Figure 3 (A), following their propagating  character for $|z|=1$, and further extended to the physically $|z|<1$ evanescent solutions.

\begin{figure}[ht]
\centering
\includegraphics[width=\columnwidth]{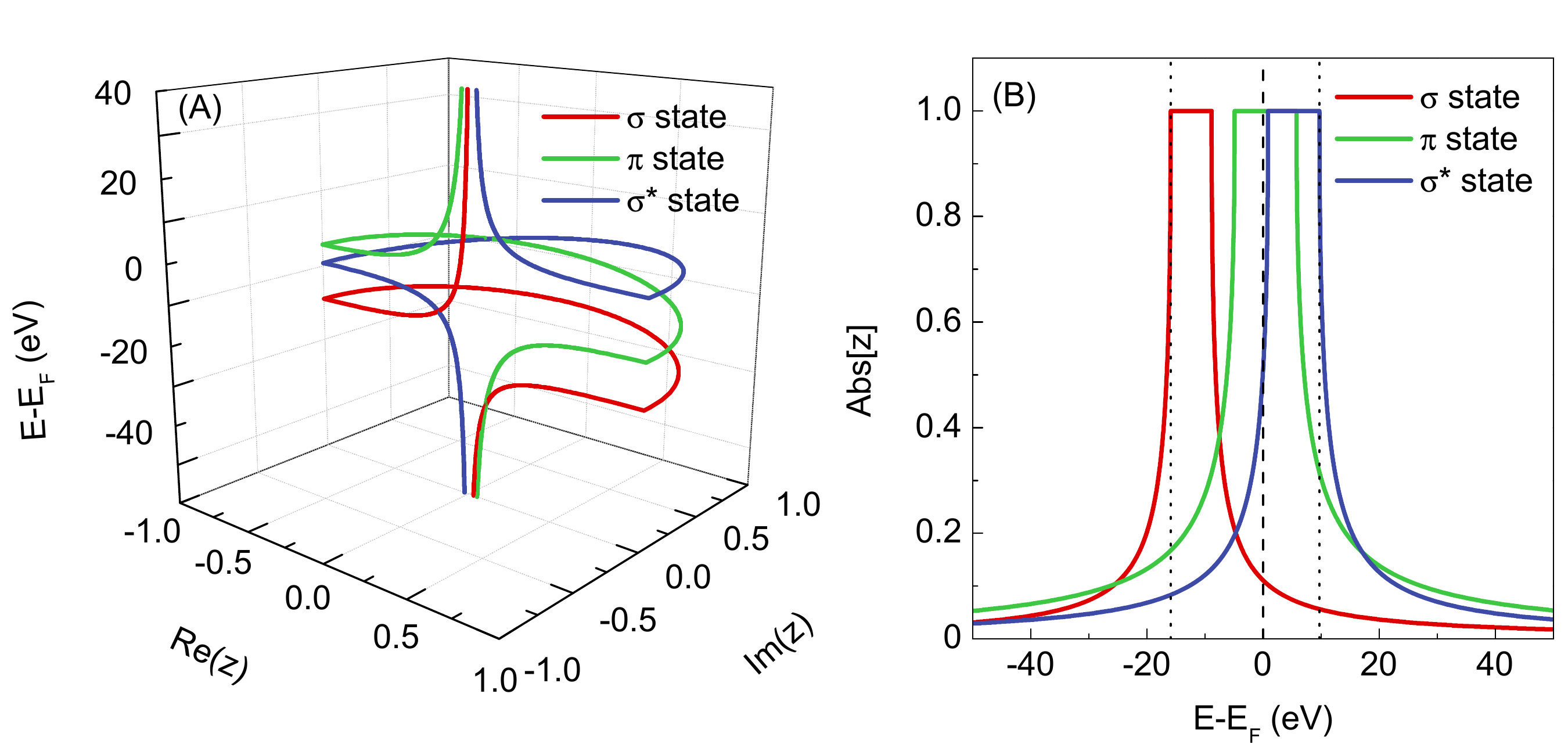}
\caption{Three-dimensional representation of the functionals $z(E)$ on a complex plane in (A), and the evolution of their absolute values as a function of energy in (B), for carbon leads. The color scheme here is the same as that for carbon in figure 3 (A).}
\label{fig02}
\end{figure}

Figures 4 provide then a more complete description for the electronic states of a given system compared to a typical band structure representation as in Figure 3, since both the propagating and evanescent states are shown. Such a general representation clearly indicates the importance of the evanescent eigenstates for a full description of the scattering problem presented in subsection \ref{transportproperties}. The energies considered in our calculations correspond to the range within the band structure boundaries, marked by two vertical dotted lines in Figure 4 (B). As a consequence not only the propagating states but also evanescent solutions are included in the quantum conductance calculations in subsection \ref{transportproperties}.

\subsection{Transport properties}
\label{transportproperties}

In this subsection the electronic transport properties of nanojunction systems composed of silicon-doped carbon wires between carbon leads, are calculated using the PFMT method. Figure 5 (A) presents a number of these systems, where we indicate the irreducible domains by the shaded grey areas. Note that these systems are always composed of finite nanojunction regions of silicon and carbon atoms, coupled to two carbon semi-infinite leads. The first three systems of Figures 5, correspond to periodic diatomic silicon carbide nanojunctions composed of respectively 1, 2, and 3 Si-C atomic pairs. The next system corresponds to a nanojunction with a substitutional disorder, composed of 3 carbon and 3 silicon atoms. The last is a symmetric nanojunction of 5 silicon atoms and only one carbon atom in the middle. Figure 5 (B) presents the group velocities of electrons in the carbon leads.

The calculated transmission and reflection scattering cross sections for each of the four available transport channels are presented in Figures 6. Each row of the figures corresponds to a nanojunction system as follows: (A)-(C) for 1,(D)-(F) for 2, (G)-(I) for 3, (J)-(L) for 4, (M)-(O) for 5. The red and green continuous curves represent the transmission and reflection spectra, respectively. The blue histograms correspond to the free electronic transport on the carbon leads, {\it i.e.} to the electronic transport on the prefect infinite quasi one-dimensional carbon wire over the different propagating states. These histograms constitute the reference to the unitarity condition which is used systematically as a check on the numerical results. The leads Fermi level is marked by a dashed line and set as a zero energy reference. Under the zero bias limit, the total conductance is calculated at this Fermi level.

\begin{figure}[ht]
\centering
\includegraphics[width=0.8\columnwidth]{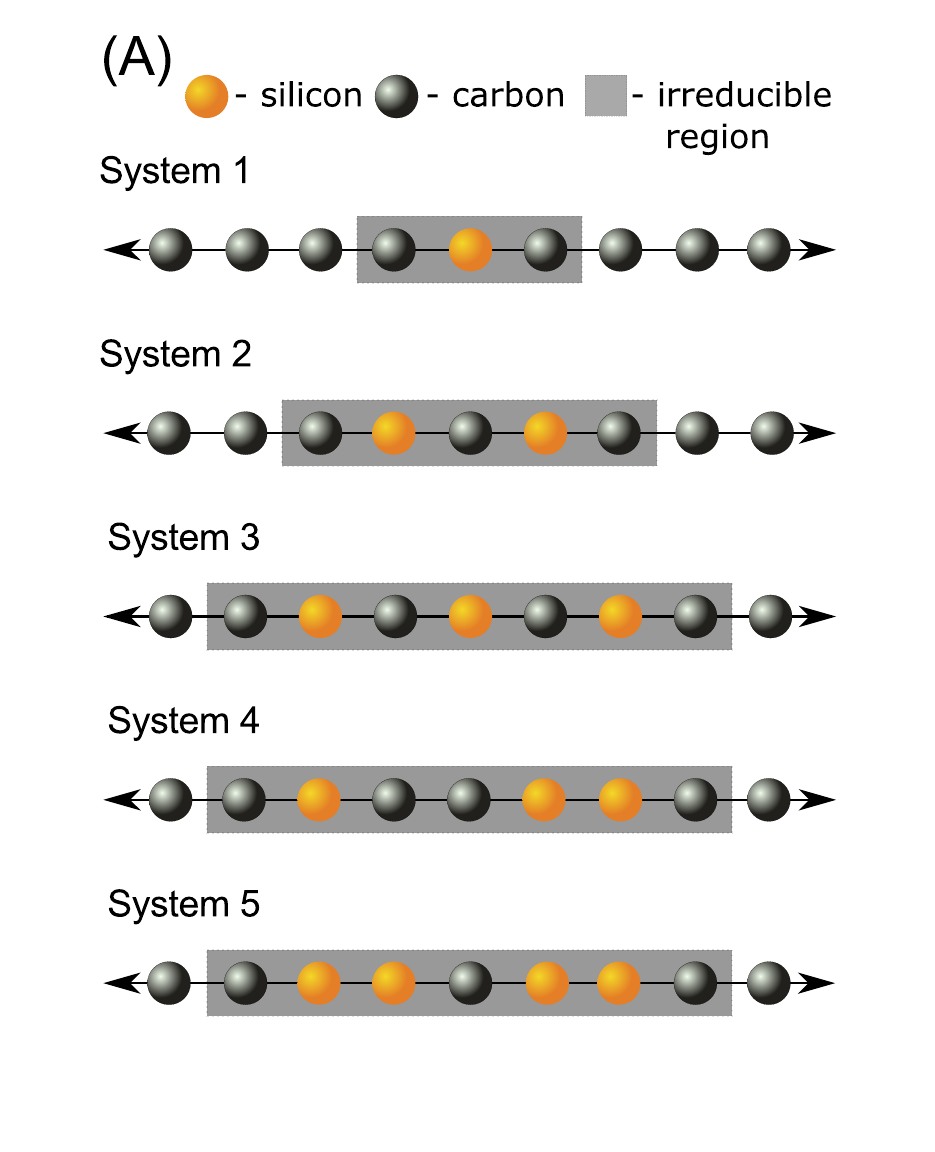}\\\includegraphics[width=0.8\columnwidth]{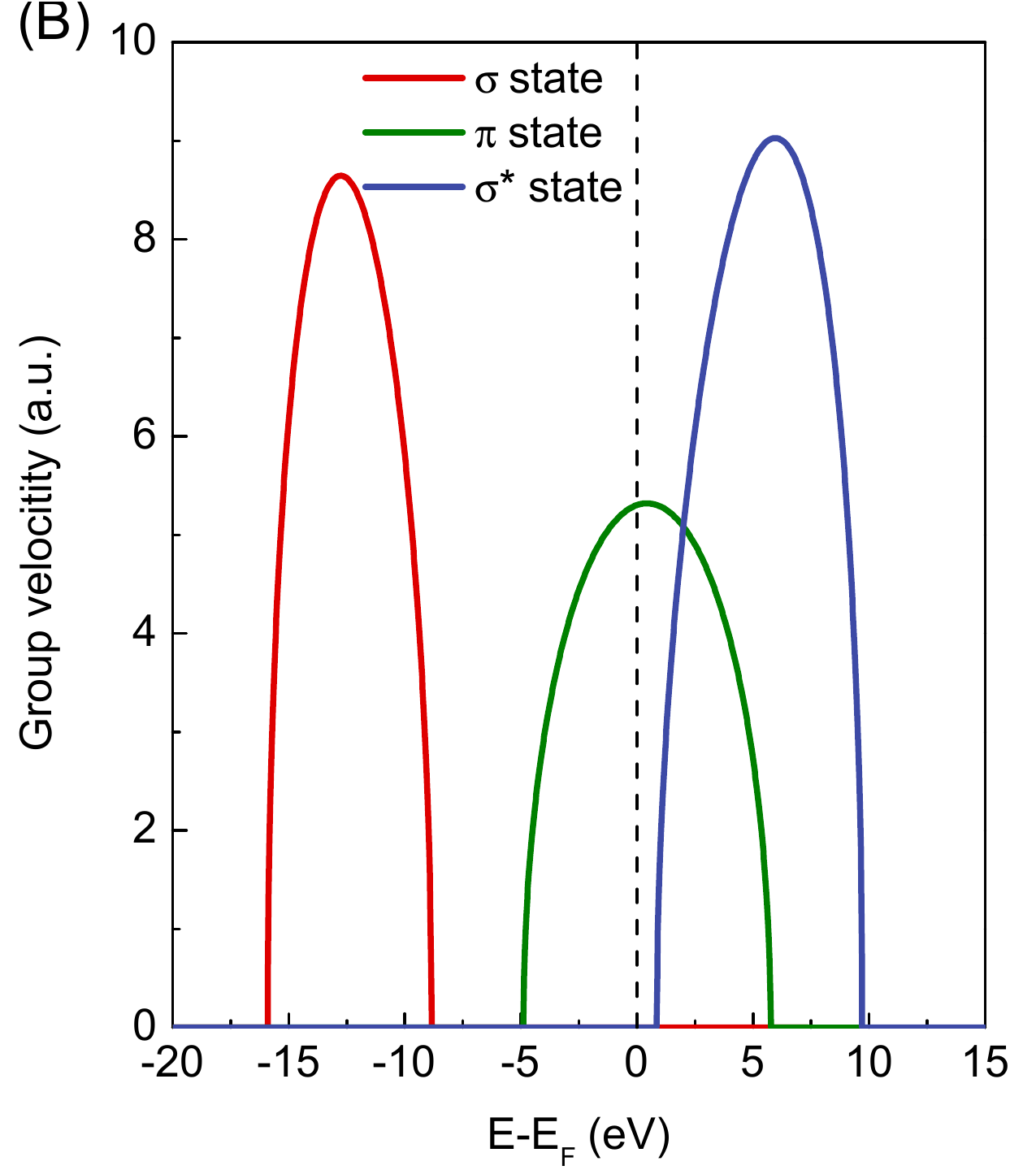}
\caption{(A) Schematic representation of the five nanojunction systems, composed of silicon and carbon atoms between one-dimensional carbon leads, considered in the present work. The irreducible domains are marked by the shaded grey areas, whereas for the other cases only the irreducible domains are shown. (B) The group velocities for the propagating band structure modes on the carbon leads.}
\label{fig02}
\end{figure}

In Figures 6, the transmission spectra present strong scattering resonances, showing increasing complexity with the increasing size and configurational order of the nanojunctions. The valence $\sigma$-state exhibits negligible transmission for all of the considered nanojunctions. The degenerate $\pi$-states and the $\sigma^{*}$-state present in contrast finite transmission spectra. However, it is only the $\pi$-states which cross the Fermi level giving rise to electronic conductance across the nanojunction within the zero bias limit.

In particular, the first three considered systems represent increasing lengths of the diatomic silicon carbide nanojunction with the increasing number of ordered Si-C atomic pairs. The transmission at the Fermi level for these systems is non-zero, see Figure 3 (C), which contrasts with the insulating character of the infinite silicon carbide wire. One can connect this finite transmission to the indirect band gap ($\Delta$) around the Fermi level for the diatomic silicon-carbide infinite wire (for more details please see Figure 3 (C)). This gap $\Delta \sim 1.5$ eV is indeed related to the difference between the binding energies of the silicon and carbon atoms, and corresponds to an effective potential barrier for the propagating $\pi$-state electrons. As the wire length increases by adding Si-C atomic pairs, as for the systems 1 to 3 of Figure 5 (B), the transmission decreases due to the cumulative barrier effects. We note that a similar effect for the monovalence diatomic copper-cobalt wire nanojunctions has been observed in a previous work \cite{szczesniak}.

\begin{figure}[ht]
\centering
\includegraphics[width=\columnwidth]{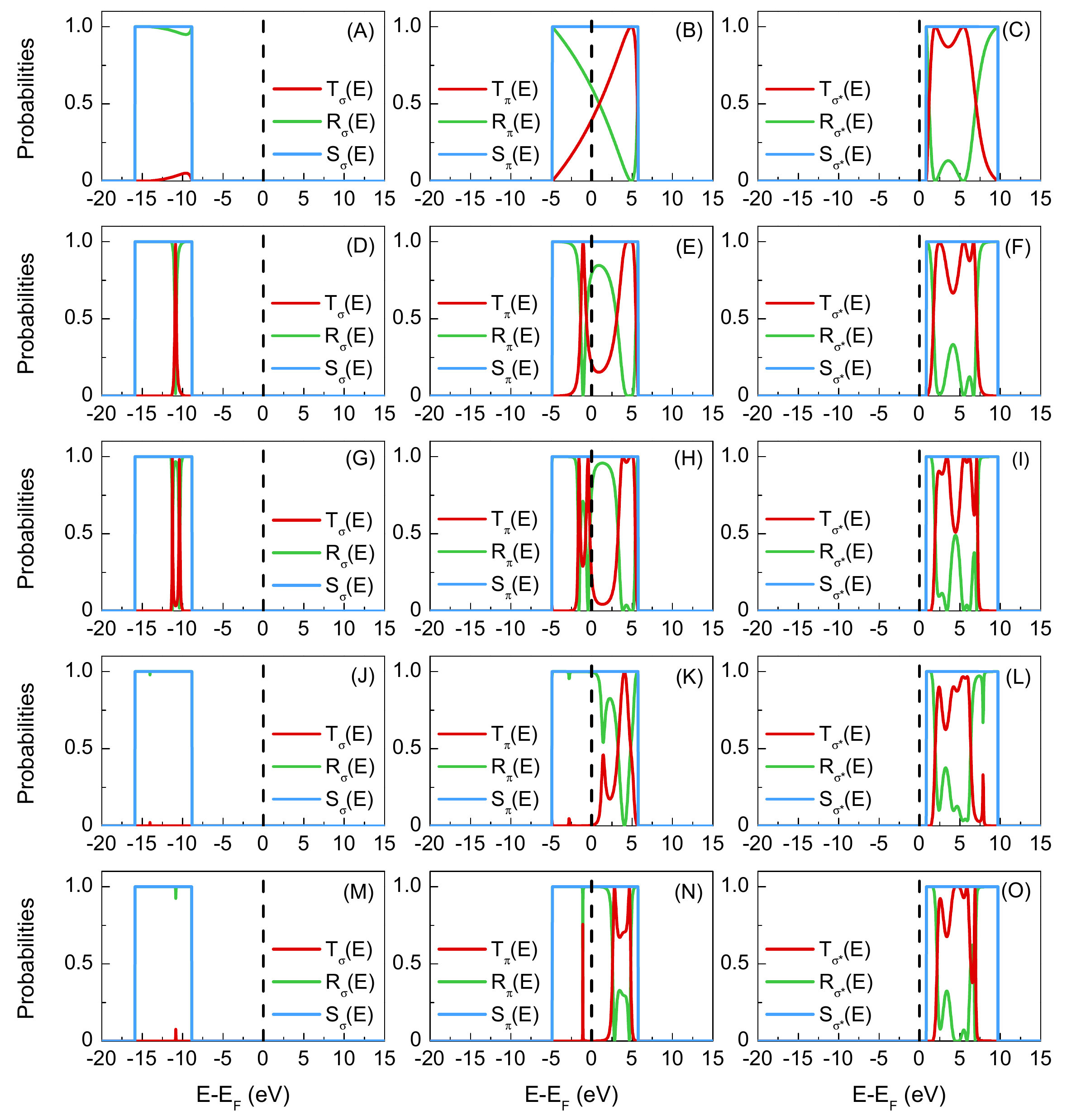}
\caption{The transmission and reflection probabilities arcross five considered types of the silicon-doped carbon wires between two semi-infinite one-dimensional carbon leads. The arrangement of the subfigures is: (A)-(C) for case 1, (D)-(F) for case 2, (G)-(I) for case 3, (J)-(L) for case 4, and (M)-(O) for case 5. The Fermi level is set at the zero energy reference position.}
\label{fig02}
\end{figure}

Furthermore, it is instructive to compare the scattering spectra for the degenerate $\pi$-states, for the nanojunction systems 3 and 4. These two systems contain identical numbers of silicon and carbon atoms, however system 3 is an ordered configuration of Si-C pairs whereas 4 presents substitutional disorder of the atoms. It is seen that the disorder suppresses the conductance of the $\pi$-state electrons at the Fermi level within the zero bias limit. Another general observation can be made from the results for the nanojunction system 5 which contains more silicon than carbon atoms. Despite the finite size of this system, comparable to system 4, and despite the structural symmetry of its atomic configuration, the electronic transmission is suppressed at the Fermi level within the zero bias limit. This implies that one of the main observations of our paper is that structural symmetry on the nanojunction is not a guarantee for finite transmission in the case of the multivalence diatomic wire nanojunctions.

Figures 6 also show that the transmission spectra for the $\sigma^{*}$-state are close to unity over a significant range of energies from $\sim$ 1 to $\sim$ 7 eV, for all of the five nanojunction systems. This result may prove useful for the electronic conductance across the silicon-doped carbon nanojunctions under finite bias voltages.

In Figures 7, we present the total electronic conductance $G(E)$ as a function of energy $E$ and in units of ${\rm G_0}=2e^2/h$, for the considered nanojunction systems of a given length as depicted in Figure 5 (red). Moreover, the perfect electronic conductance on the carbon leads (blue) is given in comparison, and constitutes effectively the conductance of the infinite and perfect quasi one-dimensional carbon wire. In Figures 7, the Fermi level is indicated by the dashed line as a zero reference energy, and $G(E)$ is calculated from all the contributing eigenstates of Figures 6, including the two degenerate $\pi$-states.

\begin{figure}[ht]
\centering
\includegraphics[width=\columnwidth]{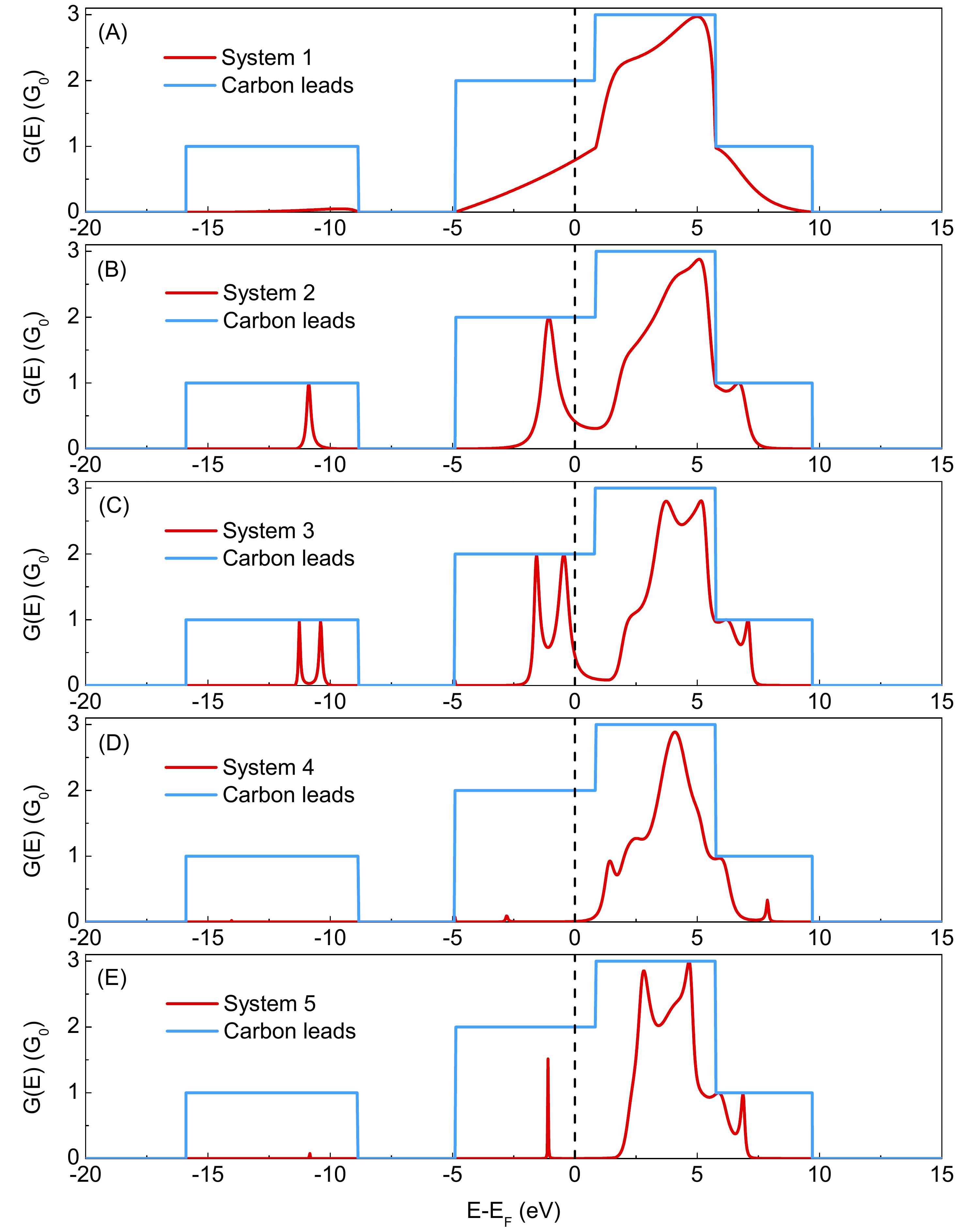}
\caption{Total electronic conductance $G(E)$, as a function of energy $E$ in units of $G_0=2e^2/h$, for silicon doped carbon wires. See text for details.}
\label{fig02}
\end{figure}

We note that the conclusions given for the results presented in Figures 6 are also followed by the more general representation of the electronic transport depicted in Figures 7. Furthermore, the results presented in Figures 7 confirm that only the electrons incident from the leads in the $\pi$ states are responsible for the electronic conductance at the zero bias limit, readable from the Fermi level position. However, for all considered systems, the conductance at the Fermi level is theoretically limited to the value of 2 ${\rm G_0}$ and the biggest conductance maxima close to the perfect infinite carbon wire value of 3 ${\rm G_0}$ can be observed only in the energy interval from $\sim$ 1 to $\sim$ 7 eV, hence for energies above the Fermi level. Once again, this follows our previous observations for the transmission results for the $\pi$ states, concluded from Figures 6. Nonetheless, only on the basis of the results presented in Figures 7 can we note that due to the summation over all possible state contributions which constitute the $G(E)$ spectra, not only the $\sigma^{*}$ state electrons but also some of those in the degenerate $\pi$ states, contribute to the high conductance values in the cited energy intervals. This important observation proves that the $\sigma^{*}$ and $\pi$ states electrons are of crucial importance for both the zero-bias quantum conductance of the silicon-doped carbon wires, and the possible finite bias ones. This implies that the use of only a single orbital for the description of the carbon atoms will result in an inadequate description of the transport processes across low-coordinated systems containing these atoms.

\section{Conclusions}
\label{conclusions}

In the present work, the unknown properties of the quantum electronic conductance for nanojunctions made of silicon-doped carbon wires between carbon leads, are studied in depth. This is done using the phase field matching theory and the tight-binding method. The local basis for the electronic wave functions is assumed to be composed of four different atomic orbitals for silicon and carbon, namely the $s$, $p_{x}$, $p_y$, and $p_z$ states.

In the first step, we calculate the electronic band structures for three nanomaterials, namely the one-dimensional infinite wires of silicon, carbon, and diatomic silicon carbide. This permits a matching comparison with the available corresponding DFT results, with the objective to select the optimal TB parameters for the three nanomaterials.

This optimal set of the tight-binding parameters is then used to calculate the electronic conductance across the silicon-doped carbon wire nanojunctions. Five different nanojunction cases are analyzed to sample their behavior under different atomic configurations. We show that despite the non conducting character of the infinite silicon carbide wires, its finite implementation as nanojunctions exhibit a finite conductance. This outcome is explained by the energy difference between the binding energies of the silicon and carbon atoms, which corresponds to an effective potential barrier for the degenerate $\pi$-state electrons transmitted across the nanojunction under zero bias field.

The conductance effects that may arise due to minimal substitutional disorder and to artificially organized symmetry  considerations on the silicon carbide wire nanojunction are also investigated.  By exchanging the positions of two silicon and carbon atoms on an initial nanojunction to generate a substitutional disordered, we show that the total quantum conductance is suppressed at the Fermi level. This is in sharp contrast with the finite and significant conductance for the initial atomically ordered nanojunction with periodic configurations of the silicon and carbon atoms. Also, the analysis of a silicon carbide nanojunction of a comparable size as the one above, presenting symmetry properties, shows that the quantum conductance is suppressed at the Fermi level.

In summary, we note that the biggest maxima of the conductance spectra {\bf for} the zero bias limit can be observed for high energies  for all of the considered systems. This conclusion reveals the fact that electrons incident from the leads in both $\sigma^{*}$ and $\pi$ states are crucial for the considerations of the electronic transport properties of the silicon-doped carbon wire nanojunctions.

\section*{Appendices}
\appendix
\section{Explicit forms of the $\mathbf{E}_{i,j}$ and $\mathbf{H}_{i,j}$ matrices}
\label{appendixa}

The explicit forms of the submatrices of equation (\ref{eq1}) are given in the following manner
\begin{equation}
\label{Aeq1}
\mathbf{E}_{i,j}=\left[ \begin{array}{c c c c c}
\boldsymbol{\varepsilon}_{1} & \mathbf{h}_{2,1}^{\dagger} & \cdots & \cdots & \mathbf{h}_{n,1}^{\dagger} \\
\mathbf{h}_{2,1} & \boldsymbol{\varepsilon}_{2} & \ddots &  & \vdots \\
\vdots & \ddots & \ddots & \ddots & \vdots \\
\vdots &  & \ddots & \boldsymbol{\varepsilon}_{n-1} & \mathbf{h}_{n,n-1}^{\dagger} \\
\mathbf{h}_{n,1} & \cdots & \cdots & \mathbf{h}_{n,n-1} & \boldsymbol{\varepsilon}_{n}
\end{array} \right],
\end{equation}
and
\begin{equation}
\label{Aeq2}
\mathbf{H}_{i,j}=\left[ \begin{array}{c c c c c c}
0 & \cdots & \mathbf{h}_{1,2} & \cdots & \mathbf{h}_{1,n-1} & \mathbf{h}_{1,n} \\
\vdots & \ddots & \ddots & \ddots & \mathbf{h}_{2,n-1} & \mathbf{h}_{2,n} \\
0 & \ddots & \ddots & \ddots & \ddots & \vdots \\
\vdots & \ddots & \ddots & \ddots & \ddots & \mathbf{h}_{n,n} \\
0 & 0 & \ddots & \ddots & \ddots & \vdots \\
0 & 0 & \cdots & 0 & \cdots & 0
\end{array} \right],
\end{equation}
where
\begin{equation}
\label{Aeq3}
\boldsymbol{\varepsilon}_{i',j'}=\left[ \begin{array}{c c c c c c}
\varepsilon_{s}^{n,\alpha} & 0 & \cdots & \cdots & 0 \\
0 & \varepsilon_{p_x}^{n,\alpha} & \ddots &  & \vdots \\
\vdots & \ddots & \ddots & \ddots & \vdots \\
\vdots &  & \ddots & \varepsilon_{l-1}^{n,\alpha} & 0 \\
0 & \cdots & \cdots & 0 & \varepsilon_{l}^{n,\alpha}
\end{array} \right],
\end{equation}
and
\begin{equation}
\label{Aeq4}
\mathbf{h}_{i',j'}=\left[ \begin{array}{c c c c c}
h_{s,s,\sigma}^{n,n',\beta} & h_{s,p_x,\sigma}^{n,n',\beta} & \cdots & \cdots & h_{s,l',m}^{n,n',\beta}  \\
h_{p_x,s,\sigma}^{n,n',\beta} & h_{p_x,p_x,\sigma}^{n,n',\beta} & \ddots &  & \vdots \\
\vdots & \ddots & \ddots & \ddots & \vdots \\
\vdots &  & \ddots & h_{l-1,l'-1,m}^{n,n',\beta} & h_{l-1,l',m}^{n,n',\beta} \\
h_{l,s,m}^{n,n',\beta} & \cdots & \cdots & h_{l,l'-1,m}^{n,n',\beta} & h_{l,l',m}^{n,n',\beta}
\end{array} \right].
\end{equation}
Equations (\ref{Aeq1}) and (\ref{Aeq2}) denote $N_x N_l$ square matrices, where matrix (\ref{Aeq2}) is upper triangular. In this manner component matrices (\ref{Aeq3}) and (\ref{Aeq4}) are of the dimension $N_l \times N_l$. Additionally, matrix $\boldsymbol{\varepsilon}_{i',j'}$ always denotes diagonal matrix, while $\mathbf{h}_{i',j'}$ matrix is much more complex, with possible non-zero elements at every position. Please note, that some of the $h_{l,l',m}^{n,n',\beta}$ elements can vanish due to the symmetry conditions and simplifies the notation of the $\mathbf{h}_{i',j'}$ matrix.

\section{Partitioning technique}
\label{appendixb}

The partitioning technique is suitable method which allows to avoid singularity problem of the $\mathbf{H}_{N,N-1}$ and $\mathbf{H}^{\dagger}_{N,N-1}$ matrices and calculate only non-trivial solutions of equation (\ref{eq10}). Detail discussion of the partitioning technique is presented in \cite{khomyakov} and this section gives only our short remarks on this method.

Following studies from \cite{khomyakov} equation (\ref{eq10}) is partitioned into two parts, of respectively $D_1-D_2$ and $D_2$ sizes, where
\begin{equation}
\label{ABeq1}
D_1 = N_x N_l,
\end{equation}
and
\begin{equation}
\label{ABeq2}
D_2 = N_n N_l.
\end{equation}
In equation (\ref{ABeq2}), parameter $N_n$ stands for the order of nearest-neighbour interactions assumed in calculations \textit{e.g.} $N_{n}=1$ for the first nearest-neighbour interactions. on the basis of equations (\ref{ABeq1}) and (\ref{ABeq2}), reduced $2N_l$ eigenvalue problem is written as
\begin{eqnarray}
\label{ABeq3}
\nonumber
&&\left[\left[ \begin{array}{c c}
\mathbf{A}_{1,1} & \mathbf{A}_{1,2} \\
\mathbf{I}_{2,2} & 0
\end{array}\right] - z
\left[ \begin{array}{c c}
\mathbf{B}_{1,1} & \mathbf{B}_{1,2} \\
0 & \mathbf{I}_{2,2}
\end{array}\right]
\right]\\
&&\times \left[ \begin{array}{c}
\mathbf{c}_{2} ( x_N, \mathbf{k} ) \\
\mathbf{c}_{2} ( x_{N-1}, \mathbf{k} )
\end{array} \right]=0.
\end{eqnarray}
At this point we correct the misprint from \cite{khomyakov} and write the submatrices of equation (\ref{ABeq3}) in the following form
\begin{equation}
\label{ABeq4}
\mathbf{A}_{1,1} = E\mathbf{I}_{2,2} - \mathbf{E}_{2,2} - \mathbf{E}_{2,1} \left[ E\mathbf{I}_{1,1} - \mathbf{E}_{1,1} \right]^{-1} \mathbf{E}_{1,2},
\end{equation}
\begin{equation}
\label{ABeq5}
\mathbf{A}_{1,2} = - \mathbf{H}_{2,2} - \mathbf{E}_{2,1} \left[ E\mathbf{I}_{1,1} - \mathbf{E}_{1,1} \right]^{-1} \mathbf{H}_{1,2},
\end{equation}
\begin{equation}
\label{ABeq6}
\mathbf{B}_{1,1} = \mathbf{H}_{2,2}^{\dagger} + \mathbf{H}_{1,2}^{\dagger} \left[ E\mathbf{I}_{1,1} - \mathbf{E}_{1,1} \right]^{-1} \mathbf{E}_{1,2},
\end{equation}
\begin{equation}
\label{ABeq7}
\mathbf{B}_{1,2} = \mathbf{H}_{1,2} \left[ E\mathbf{I}_{1,1} - \mathbf{E}_{1,1} \right]^{-1} \mathbf{H}_{1,2}
\end{equation}
Please note, that reduced problem of equation (\ref{ABeq3}) gives $2N_l$ eigenvalues with $2N_l$ corresponding eigenvectors, what is $N_x$ times less then can be expect from the physical point of view. Nevertheless, those solutions can be easily separated into $N_x N_l$ eigenvalues and $N_x N_l$ eigenvectors of a purely physical character.
\section{Explicit forms of the $\mathbf{M}_{i,j}$, $\mathbf{M}_{1}^{in}$, and $\mathbf{M}_{2}^{in}$ components}
\label{appendixc}

The submatrices of the {\it matched} $(D+2) \times (D+2)$ square matrix $\mathbf{M}$ in equation (\ref{eq16}), for a given $i$ and $j$ indices, are given as
\begin{equation}
\label{ACeq1}
\mathbf{M}_{i,j} = E\mathbf{I}-\mathbf{E}_{i,i} \hspace{0.2cm} \textrm{for} \hspace{0.2cm}
\left\{ \begin{array}{c c}
i=j\\
D>i>1\\
D>j>1
\end{array} \right. ,
\end{equation}
\begin{equation}
\label{ACeq2}
\mathbf{M}_{i,j} = -\mathbf{H}_{i,i-1} \hspace{0.2cm} \textrm{for} \hspace{0.2cm}
\left\{ \begin{array}{c c}
i \neq j\\
i>2\\
j=i-1
\end{array} \right. ,
\end{equation}
\begin{equation}
\label{ACeq3}
\mathbf{M}_{i,j} = -\mathbf{H}^{\dagger}_{i,i-1} \hspace{0.2cm} \textrm{for} \hspace{0.2cm}
\left\{ \begin{array}{c c}
i \neq j\\
i<D+1\\
j=i+1
\end{array} \right. ,
\end{equation}
except of the submatrices which describe the boundary atoms of the system and are expressed in the following manner
\begin{eqnarray}
\label{ACeq4}
\nonumber
&&\mathbf{M}_{1, 1} = -\mathbf{H}_{-1,-2} \mathbf{c}_{l} (\mathbf{r}_{n},z_{\gamma'}, E_\gamma)z_{\gamma'}^{2}\\
\nonumber
&&+ \left( E\mathbf{I} - \mathbf{E}_{-1,-1} \right) \mathbf{c}_{l} (\mathbf{r}_{n},z_{\gamma'}, E_\gamma)z_{\gamma'},\\
\nonumber
&&\mathbf{M}_{2, 1} = -\mathbf{H}_{0,-1} \mathbf{c}_{l} (\mathbf{r}_{n},z_{\gamma'}, E_\gamma)z_{\gamma'},\\
\nonumber
&&\mathbf{M}_{D+1, D+2} = -\mathbf{H}^{\dagger}_{D-1,D} \mathbf{c}_{l} (\mathbf{r}_{n},z_{\gamma'}, E_\gamma)z_{\gamma'}^{D},\\
\nonumber
&&\mathbf{M}_{D+2, D+2} = -\mathbf{H}_{D,D+1} \mathbf{c}_{l} (\mathbf{r}_{n},z_{\gamma'}, E_\gamma)z_{\gamma'}^{D+1}\\
&&+ \left( E\mathbf{I} - \mathbf{E}_{D,D} \right) \mathbf{c}_{l} (\mathbf{r}_{n},z_{\gamma'}, E_\gamma)z_{\gamma'}^{D},.
\end{eqnarray}
Finally, the $\mathbf{M}^{in}_{1}$ and $\mathbf{M}^{in}_{2}$ of equation (\ref{eq16}) vector components are written as
\begin{eqnarray}
\label{ACeq8}
\nonumber
\mathbf{M}^{in}_{1} = \mathbf{H}_{-1,-2} \mathbf{c}_{l} (\mathbf{r}_{n}, z_\gamma, E_\gamma) z_{\gamma}^{-2}\\
+ \left( E\mathbf{I} - \mathbf{E}_{-1,-1} \right) \mathbf{c}_{l} (\mathbf{r}_{n}, z_\gamma, E_\gamma) z_{\gamma}^{-1},
\end{eqnarray}
and
\begin{equation}
\label{ACeq9}
\mathbf{M}^{in}_{2} = -\mathbf{H}_{0,-1} \mathbf{c}_{l} (\mathbf{r}_{n}, z_\gamma, E_\gamma) z^{-1}.
\end{equation}
%
\section{Group velocities}
\label{appendixd}
As specified in section \ref{pfmt}, the group velocities for individual states, can be calculated on the basis of equation (\ref{eq6}) rewritten in the following manner
\begin{equation}
\label{ADeq1}
\left[ v \mathbf{I} - \mathbf{V} \right] \mathbf{v}(\mathbf{k}, E)=0,
\end{equation}
where $v$ denotes the eigenvalues of equation (\ref{ADeq1}) which yields all required electron group velocities for each propagating state. Further $\mathbf{V}$ is the $N_x \times N_l$ size matrix of the form
\begin{equation}
\label{ADeq2}
\mathbf{V}=\frac{\partial\mathbf{M}_d}{\partial\mathbf{k}}.
\end{equation}
Finally $\mathbf{v}(\mathbf{R}_N, \mathbf{k})$ stands for eigenvectors of the problem of equation (\ref{ADeq1}). We note that usually equation (\ref{ADeq2}) includes the constant part $d_{\beta}/h$, where $h$ is the Planck constant. However, for the purpose of electronic conductance calculations within the PFMT approach, this term can be omitted due to the fact that only the ratios of the given group velocities are important (please see equations (\ref{eq17}) and (\ref{eq18})).

\section*{Author's contributions}
DS participated in the design of this study, analytical calculations and writing the code for numerical calculations, carried out numerical calculations, drafted the manuscript, and participated in writing the final version of the manuscript. AK coordinated this study, participated in its design and its  analytical calculations, and in writing the final version of the manuscript. ZB participated in the design of this study, its coordination, and substantial critical revision of the final version of the manuscript. RS participated in writing the code for numerical calculations, and substantial critical revision of the final version of the manuscript. MG participated in substantial critical revision of the final version of the manuscript. All authors read and approved the final manuscript.


\section*{Competing interests}
The authors declare that they have no competing interests.


\section*{Acknowledgements}

D. Szcz{\c e}{\' s}niak, would like to thank the French Ministry of Foreign Affairs for his Ph.D. scholarship grant CNOUS 2009-2374, also the Polish National Science Center for their research grant DEC- 2011/01/N/ST3/04492, and the Graduate School of Sciences at the University du Maine for its support.

\bibliographystyle{apsrev}
\bibliography{manuscript}
\end{document}